%% file: main.tex
\documentclass{Interspeech}

\usepackage{tipa}
\usepackage[utf8]{inputenc}
\usepackage[T1]{fontenc}
\usepackage{graphicx}
\usepackage{booktabs}   
\usepackage{color}
\usepackage[table,xcdraw]{xcolor} 

\definecolor{brightturquoise}{rgb}{0.85, 1, 1}

\interspeechcameraready 

\makeatletter
\renewcommand\anonaddress{%
  \ifinterspeechfinal
    {\centering
      \vskip-3pt
      \shortstack[c]{%
        \textsuperscript{1}Zhejiang University, China\quad
        \textsuperscript{2}UC Berkeley, United States\\
        \textsuperscript{3}UCSF, United States
      }\par
      \vskip-3pt
    }%
  \fi
}
\makeatother

\title{Towards Accurate Phonetic Error Detection Through Phoneme Similarity Modeling}

\author[affiliation={ 1 }]{Xuanru}{Zhou}
\author[affiliation={ 2 }]{Jiachen}{Lian$^*$}
\author[affiliation={ 2 }]{Cheol Jun}{Cho}
\author[affiliation={ 2 }]{Tejas}{Prabhune}
\author[affiliation={ 1 }]{Shuhe}{Li}
\author[affiliation={ 2 }]{William}{Li}
\author[affiliation={ 2 }]{Rodrigo}{Ortiz}
\author[affiliation={ 3 }]{Zoe}{Ezzes}
\author[affiliation={ 3 }]{Jet}{Vonk}
\author[affiliation={ 3 }]{Brittany}{Morin}
\author[affiliation={ 3 }]{Rian}{Bogley}
\author[affiliation={ 3 }]{Lisa}{Wauters}
\author[affiliation={ 3 }]{Zachary}{Miller}
\author[affiliation={ 3 }]{Maria}{Gorno-Tempini}
\author[affiliation={ 2 }]{Gopala}{Anumanchipalli}

\affiliation{}{Zhejiang University}{China}
\affiliation{}{UC Berkeley}{US}
\affiliation{}{UCSF}{US}
\email{$^*$ Project Lead, corresponding to: jiachenlian@berkeley.edu}


\usepackage{comment}

\begin{document}

\maketitle

\input{abstract}
\input{introduction}
\input{method}
\input{experiments}
\input{conclusion}

\bibliographystyle{IEEEtran}
\bibliography{mybib}

\end{document}

%% file: abstract.tex
\begin{abstract}
% 1000 characters. No citations.
Phonetic error detection, a core subtask of automatic pronunciation assessment, identifies pronunciation deviations at the phoneme level. Speech variability from accents and dysfluencies challenges accurate phoneme recognition, with current models failing to capture these discrepancies effectively. We propose a verbatim phoneme recognition framework using multi-task training with novel phoneme similarity modeling that transcribes what speakers actually say rather than what they're supposed to say. We develop and open-source \textit{VCTK-accent}, a simulated dataset containing phonetic errors, and propose two novel metrics for assessing pronunciation differences. Our work establishes a new benchmark for phonetic error detection.

\end{abstract}

%% file: introduction.tex
\section{Introduction}

Speech pronunciation assessment plays a crucial role in language learning~\cite{kheir-etal-2023-automatic, truong04_icall} and diagnosis of speech disorders~\cite{ssdm}. As traditional human assessment is time-consuming and lacks unified standards, recent advancement has shifted to Computer-Aided Pronunciation Training (CAPT)~\cite{lee2016language-capt}.
Parallel to the bloom of textual large language models~\cite{OpenAI_chatgpt}, recent speech research focuses on audio-language foundation models~\cite{ji2024wavchat, huang2024dynamic} that unify multiple tasks including pronunciation assessment. However, little evidence shows that multi-task learning improves performance in specific tasks like pronunciation modeling. Thus, \textit{automatic pronunciation assessment remains a task-specific area} requiring domain-specific priors in model design.

\begin{figure}[t]
    \centering
    \includegraphics[height=4.8cm]{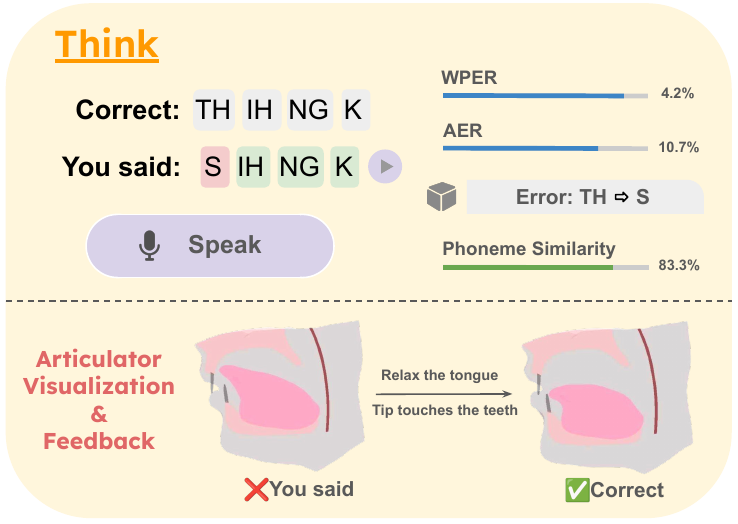}
    \vspace{-5pt}
    \caption{Demo of our method. Model transcribes user speech into phoneme sequences, detects errors, scores using metrics, and generates articulator visualizations and feedback~\cite{cho2024jstsp}: relax the tongue and place it against the roof of the mouth, with the tip lightly touching the teeth.}
    \label{fig:demo}
\end{figure}

Automatic pronunciation assessment involves detecting multiple aspects of speech quality~\cite{lee2016language-capt}, including fluency, intonation, phonetic errors, prosodic features, stress patterns, and rhythmic consistency. We focus on a core module: phonetic error detection, which aims to identify pronunciation deviations at a fine-grained phoneme level. This is fundamentally an \textit{allophony} problem~\cite{boomershine2008impact-allophony}. Specifically, if a person pronounces phoneme A, how confidently can we determine its relation to phoneme $\hat{A}$? For instance, can AI models reliably distinguish between your pronunciation of the phoneme ``TH'' when the ground truth is ``S'' in the word \textit{think}, as illustrated in Figure~\ref{fig:demo}? So far, we have found no evidence of such capabilities in mainstream language learning platforms such as Duolingo~\cite{Duolingo}, Speak~\cite{SpeakApp}, or even GPT-4o Voice~\cite{OpenAI_GPT4o}. 

Earlier studies have treated pronunciation allophony as a phoneme recognition or classification task, using normalized classification logits scores as confidence measures~\cite{ssdm,yang22v_interspeech, lian2023unconstrained-udm, lian-anumanchipalli-2024-towards-hudm,lian2024ssdm2.0}. These are also called phonetic replacement errors in dysfluency modeling~\cite{lian2023unconstrained-udm, lian-anumanchipalli-2024-towards-hudm, zhou24eyolostutter, zhou2024stutter-solver, zhou2024timetokensbenchmarkingendtoend, ssdm, lian2024ssdm2.0,zhang2025analysisevaluationsyntheticdata, guo2025dysfluentwfstframeworkzeroshot, ye2025seamlessalignment-neurallcs, 11036667}. While~\cite{lee2025dypcl} explored phoneme similarity in CTC training for dysarthric speech, their approach focuses on transcribing what speakers should have said rather than what they actually said, which is our research focus. Another relevant work~\cite{choi2025leveragingallophonyselfsupervisedspeech} demonstrated that self-supervised speech representations can implicitly leverage phoneme similarity in a way that aligns more closely with human perception. However, it primarily focused on analysis and still fails to transcribe what the person actually said. Our work, on the other hand, focuses on the latter task, which is \textit{distinct} and more \textit{challenging}.

In this paper, we propose a framework for verbatim phoneme recognition with phoneme similarity modeling, trained using multi-task learning~\cite{crawshaw2020multitasklearningdeepneural}: phoneme mapping and connectionist
temporal classification (CTC), aiming to accurately transcribe the phonemes pronounced by the speaker.
Phoneme similarity modeling, which serves as a method to reveal allophony information, better aligns with both how humans produce speech and how the human ear perceives it. We thus propose three novel phoneme similarity modeling methods: heuristic-based, articulatory-based, and Sylber-based~\cite{cho2024sylber} methods. Through this modeling, we compute the similarity score between each pair of phonemes. The process of integrating phoneme similarity into the training procedure is referred to as \textit{soft-training}.
To providing training material that incorporates phonetic errors and accurate labels, we follow the TTS-based simulation approach~\cite{zhou24eyolostutter} and generate a simulated dataset with vowel and consonant substitutions, which we named \textit{VCTK-accent} and open-sourced. In addition, We introduce two novel metrics: \textit{WPER} and \textit{AER}, which can be used not only for evaluating phoneme recognition models but also as pronunciation assessment scores for speech.

Evaluated on the VCTK-accent test sets and real speech datasets, our method shows impressive performance. Our training strategy-\textit{multi-task learning \& soft-training} significantly reduced the PER, WPER, and AER, proving its effectiveness. This highlights the crucial role of phoneme similarity modeling as a key approach to tackling the problem. Our work establishes a new benchmark for phonetic error detection. The project page is available
at \url{https://berkeley-speech-group.github.io/Phonetic-Error-Detection/}.

%% file: method.tex
\section{Method}

\subsection{Data Simulation}
\label{sec:data-simulation}
To train the phoneme recognition model, accurate labels are essential, mapping directly to the word being pronounced. Thus, we follow the TTS-based dysfluency simulation pipeline described in~\cite{zhou24eyolostutter}. First, we inject phonetic substitutions into each word from the text of the VCTK corpus~\cite{yamagishi2019cstr-vctk} based on a predefined set of common phoneme substitution pairs listed in Table. ~\ref{table-sub}, which include vowel-to-vowel and consonant-to-consonant substitutions respectively. Next, we input the modified IPA sequences into the VITS~\cite{kim2021conditional-vits} model to generate speech with phonetic errors. These resulting \textit{(speech, modified phoneme sequence, reference word)} pairs are used as training data for our phoneme recognition model.

\vspace{-5pt}
\begin{table}[h]
    \caption{Common CMU phoneme substitution pairs}
    \label{table-sub}
    \centering
    \setlength{\tabcolsep}{6pt} 
    \renewcommand{\arraystretch}{1.2} 
    \resizebox{8cm}{!}{
    \begin{tabular}{c c c | c c c c} 
     \toprule
    \multicolumn{3}{c|}{\textbf{Vowel}} & \multicolumn{4}{c}{\textbf{Consonant}} \\
    \hline
    \hline
    % \textipa{(a, i)} & \textipa{(æ, u)} & \textipa{(\char"0250, \char"026A)} & \textipa{(p, ɡ)} & \textipa{(t, ʒ)} & \textipa{(k, b)} & \textipa{(m, s)} \\
    % \textipa{(o, ɛ)} & \textipa{(ɔ, e)} & \textipa{(ʊ, ɜ)} & \textipa{(n, ʃ)} & \textipa{(ŋ, f)} & \textipa{(l, t)} & \textipa{(ɹ, d)} \\
    % \textipa{(ə, i)} & \textipa{(ɚ, o)} & \textipa{(ʌ, æ)} & \textipa{(w, k)} & \textipa{(θ, v)} & \textipa{(ð, z)} & \textipa{(ʃ, h)} \\
    (AA, IY) & (AE, UW) & (AA, IH) & (P, G) & (T, ZH) & (K, B) & (M, S) \\
    (OW, EH) & (AO, EH) & (UH, ER) & (N, SH) & (NG, F) & (L, T) & (R, D) \\
    (AH, IY) & (ER, OW) & (AH, AE) & (W, K) & (TH, V) & (DH, Z) & (SH, HH) \\
    \bottomrule
    \end{tabular}}
\end{table}

\vspace{-5pt}
\subsection{Phoneme Similarity Modeling} \label{sec:phn-similarity}
Unlike~\cite{lee2025dypcl} directly obtained the phoneme distance using PanPhon tool~\cite{mortensen-etal-2016-panphon}, we propose three methods for modeling phoneme similarity: heuristic-based, articulatory-based and Sylber-based methods. By incorporating perspectives from acoustics and syllables, these approaches more closely align with human pronunciation and auditory perception. We obtain a phoneme similarity matrix $S \in \mathbb{R}^{N \times N}$, where each value in the range $(0, 1)$, indicating the similarity between each pair of phonemes. $N$ represents the size of the phoneme dictionary.

\subsubsection{Heuristic-based}
\begin{figure}[t]
    \centering
    \includegraphics[height=6.8cm]{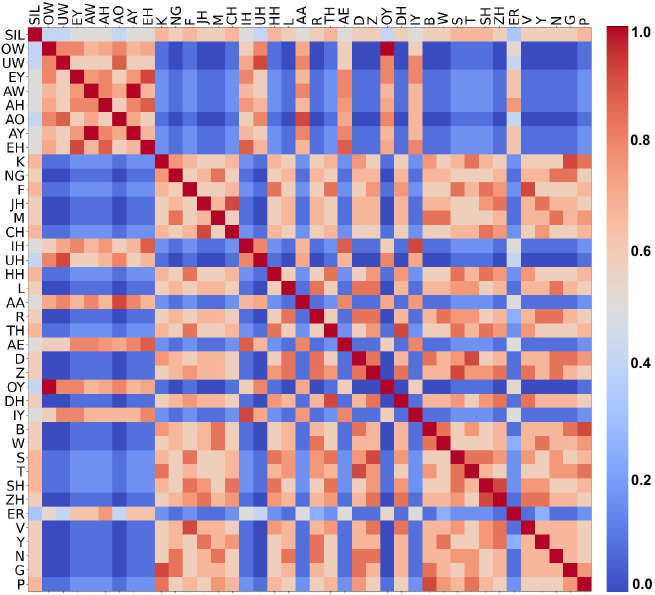}
    \caption{Heuristic-based phoneme similarity matrix}
    \label{fig:sim}
\end{figure}

Heuristic-based method calculates similarity by comparing phonemes based on eight features~\cite{anderson2018essentials}: vowel or consonant, vowel length, vowel height, vowel frontness, lip rounding, consonant type, place of articulation, and consonant voicing. Each feature is assigned a normalized weight, and the similarity score between pairs of phonemes is computed by summing the weights of matching feature values. In this study, the weights are set as follows: 0.2, 0.1, 0.15, 0.15, 0.1, 0.2, 0.2, and 0.1, respectively. The visualization of the phoneme similarity matrix is presented in Figure~\ref{fig:sim}.

\subsubsection{Articulatory-based} \label{sec:art-based}
We first construct a reference articulatory position for each phoneme using the VCTK corpus and an acoustic-to-articulatory inversion (AAI) model~\cite{cho2024jstsp}, employing a data-driven approach. Next, we compute the L2 distance between each pair of phonemes and apply min-max normalization to obtain the similarity scores.

\subsubsection{Sylber-based}
Human speech segmentation is natually syllabic~\cite{cho2024sylber}. We then utilize the Sylber~\cite{cho2024sylber} feature, as it offers a clean and robust syllabic structure. First, we fine-tune Sylber using a phoneme classification task with a single linear classifier layer, employing the VCTK corpus. Then, following a similar approach outlined in Sec.~\ref{sec:art-based}, we construct a reference feature for each phoneme and compute the similarity score.

Overall, the heuristic-based method stems from the perspective of phoneme classification and definition. Both the articulatory and Sylber-based methods are data-driven approaches: the former emphasizes the acoustic aspect, while the latter focuses more on the syllabic aspect.

\subsection{Verbatim Phoneme Recognition}
\begin{figure*}[ht]
    \centering
    \includegraphics[width=16.5cm]{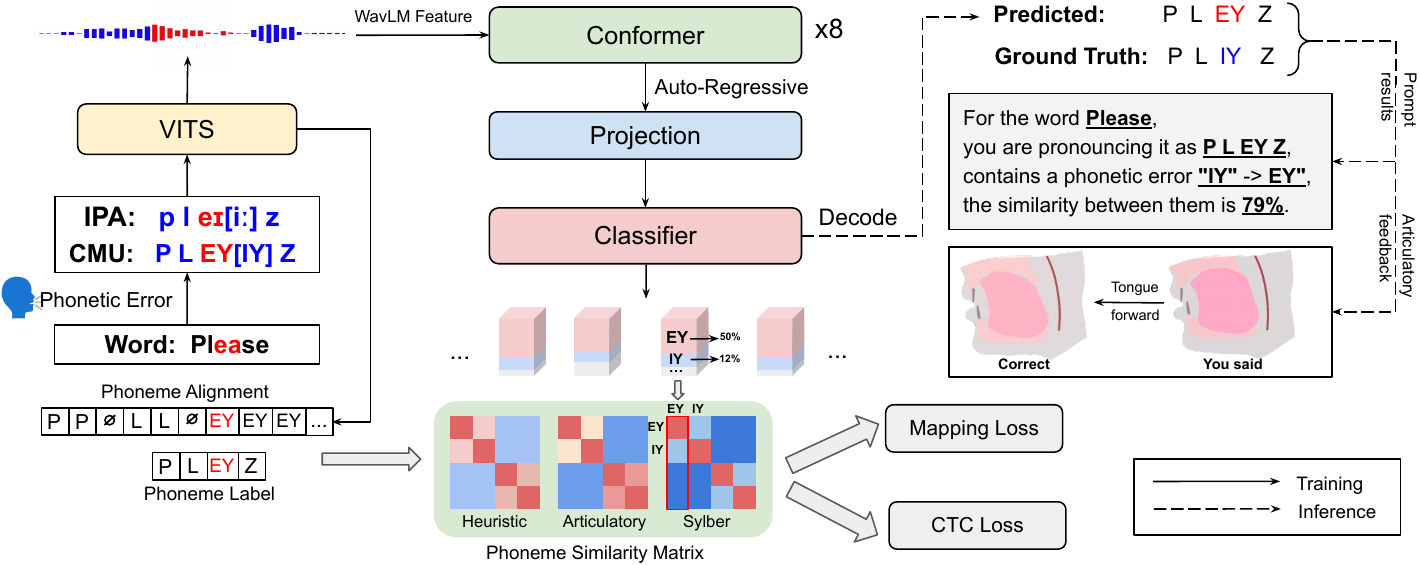}
    \caption{Pipeline of phoneme recognition and error detection: phonetic error of "IY" -> "EY" with a similarity score 79\% in the word "Please", and the articulatory feedback is moving the tongue towards the front of the mouth.}
    \label{fig:pipeline}
\end{figure*}

We adopt the speech feature extracted from WavLM~\cite{wavlm} to train a verbatim phoneme recognition model with multi-task learning and soft-training. 
The entire pipeline is illustrated in Figure~\ref{fig:pipeline}, and the model architecture and training objectives are described in the following sections.

\subsubsection{Model}
The phoneme recognition model consists of a conformer~\cite{conformer} encoder, an autoregressive projection layer, and a linear classifier. This conformer encoder consists of 8 layers with 4 attention heads per layer. For the auto-regressive mechanism, For the autoregressive mechanism, at each timestep, the model combines the output features from the conformer encoder with the embedding of the previously predicted phoneme to predict the next phoneme, thereby effectively capturing the sequential dependencies in the data.

\subsubsection{Multi-task Learning and Soft-training}
We employ weighted loss-based multi-task training pipeline with two separate loss values: soft-CTC loss and soft-mapping loss. 
For the phoneme similarity matrix, each column(row) represents the similarity score of this phoneme and all other phonemes. We can treat this vector as a \textit{soft label} of this phoneme.
For soft-CTC loss, we utilize the target phoneme's soft label to weight the emission probability, thereby reducing the penalty for prediction errors between similar phonemes.
Traditional cross-entropy loss treats all phonemes as independent and ignores their similarities. Thus, in soft-mapping loss, we replace the one-hot encoded target with the target phoneme's soft label.
The complete loss is shown below:

\vspace{-5pt}
\begin{align} \label{equation:multi-loss}
\mathbb{L} &= \lambda_{ctc} \cdot L_{sCTC} + \lambda_{map} \cdot L_{smap} \notag \\
&= -\lambda_{ctc} \cdot \log \left( \sum_{z \in \mathcal{B}(y)} \prod_{t=1}^{T} \sum_{j=1}^{N} \hat{S}(z_t, j) \cdot p_t(j) \right) \notag \\
&\quad +\lambda_{map} \cdot \sum_{t=1}^{T} \sum_{j=1}^{N} (p_{t}(j) - \hat{S}(y_t,j))^2
\end{align}
Where $S$ denotes the normalized phoneme similarity matrix, $\mathcal{B}(y)$ denotes the set of all compatible alignments, $y$ is the target phoneme sequence, $p$ is model's output emission probability, and and $N$ is the phoneme dictionary size. In this work, we set $\lambda_{CTC}=0.8$, $\lambda_{map}=0.2$.

% \subsubsection{Contrastive Learning}

% \begin{align} \label{equation:contrasive}
%    \mathbb{L} = (\frac{1}{3}\sum^{(A,P,N)} L_{CTC} + L_{smap}) + \lambda \cdot L_{triplet}(a,\ p,\ n)
% \end{align}

% where $A$, $P$, and $N$ denote the anchor, positive,
% and negative audio samples, respectively, and $a$,
% $p$, and $n$ are the corresponding anchor, positive,
% and negative phonemes.

%% file: experiments.tex
\section{Experiment}

\subsection{Datasets}
\begin{itemize}

\item \textbf{VCTK-Accent} is a TTS-based~\cite{zhou24eyolostutter} simulated datasets, extended from VCTK corpus~\cite{yamagishi2019cstr-vctk}, which contains vowel and consonant phonetic errors, with simulation details provided in Sec.~\ref{sec:data-simulation}. The total duration of the dataset is 323.9 hours.

\item \textbf{L2-ARCTIC~\cite{zhao18L2Arctic}} 
includes recordings from 24 non-native English speakers, each recording about one hour of read speech from CMU’s ARCTIC prompts. It provided forced-aligned phonetic transcriptions, and annotated 150 utterances per speaker for mispronunciation errors.

\item \textbf{Speechocean762 ~\cite{zhang21speechocean}} is an open-source non-native English speech corpus, which consists of 5000 English sentences. All the speakers are non-native, and their mother tongue is Mandarin, and half of the speakers are children.

\item \textbf{MultiPA~\cite{chen24multipa}} was collected from real-world open-response scenarios and consists of 50 audio clips, each lasting 10 to 20 seconds, from around 20 anonymous Mandarin-speaking users practicing English with a dialog chatbot.

\item \textbf{PPA Speech} is collected from our clinical collaborators, and consists of recordings from 38 participants diagnosed with Primary Progressive Aphasia (PPA)~\cite{gorno2011classification-ppa}. We selected segments containing phonetic errors for evaluation.
% People were asked to read grandfather passage, leading about \textcolor{red}{X} hour of speech.
\end{itemize}
For the above real speech datasets, we did the segmentation and annotation ourselves, the process is: we first segment out the words with phonetic error, then label the phoneme sequence that the speaker actually pronounced as the target.

\vspace{-5pt}
\subsection{Training}
The phoneme recognition model is trained with 90/10 train/test split on VCTK-accent, with a batch size of 256. We use Adam optimization and gradient clipping, and the learning rate is 3e-4. The model is trained for 30 epochs with total of 75 hours on an RTX A6000.

\subsection{Evaluation Metrics}
\vspace{-3pt}
\subsubsection{Phoneme Error Rate (PER)}
PER measure of how many errors (inserted, deleted, and substitute phonemes) are predicting phoneme sequences compared to the actual phoneme sequence. It calculated by dividing the number of phoneme errors by the total number of phonemes.

\vspace{-5pt}
\subsubsection{Weight Phoneme Error Rate (WPER)}
The single-word phoneme error rate (PER) is a relatively coarse metric, as it only reflects the presence or absence of phonetic errors, lacking the ability to capture nuanced pronunciation differences. To address this limitation, we introduce a more refined metric: the Weighted Phoneme Error Rate (WPER). In the case of substitutions, we replace the count of substitutions with the sum of the phoneme similarities between the substituted pairs. The equation is shown below:

\begin{table*}[htp]
    \caption{Evaluation on VCTK-accent testsets with three different phoneme similarity modeling methods}
    \label{table-eval-vctk}
    \centering
    \setlength{\tabcolsep}{7pt} 
    \renewcommand{\arraystretch}{1.1} 
    \resizebox{15cm}{!}{
    \small
    \begin{tabular}{l|c c c|c c c|c c c} 
     \toprule
      Metrics& \multicolumn{3}{c|}{Vowel}& \multicolumn{3}{c|}{Consonant} & \multicolumn{3}{c}{All}\\
      \rowcolor{brightturquoise}
     & \textit{Heuristic} & \textit{Articulatory} & \textit{Sylber} & \textit{Heuristic} & \textit{Articulatory} & \textit{Sylber} & \textit{Heuristic} & \textit{Articulatory} & \textit{Sylber}\\
     \hline
    \hline
    PER ($\%$, $\downarrow$) & 12.39 & 10.15 & 10.04 & 13.93 & 13.62 & 12.89 & 15.85 & \textbf{12.37} & 16.51\\
    WPER ($\%$, $\downarrow$)& 8.19 & 7.34 &8.83 & 9.18 & 7.75 & 10.42 & 9.39 & \textbf{7.41} & 9.57\\
    AER ($\%$, $\downarrow$) & 9.22 &7.93 & 9.37 & 10.84 & 10.76 & 11.09 & 12.82 & \textbf{10.53} & 12.08 \\
    \bottomrule
    \end{tabular}}
\end{table*}
\vspace{-8mm}
\begin{align}
   WPER = \frac{D + \displaystyle \sum^{(p_r, p_s)} (1 - S(p_r, p_s)) + I}{L}
  \label{equation:wper}
\end{align}
Where $L$ be the length of the reference phoneme sequence, $D$ and $I$ the counts of deletions and insertions, $S$ the phoneme similarity matrix constructed in Sec.~\ref{sec:phn-similarity}, $p_r$, $p_s$ are the reference and substitute phoneme, respectively.

\subsubsection{Articulatory Error Rate (AER)}
We also propose a metric that considers the articulatory distance between different phonemes. For each speech sample, we apply the AAI model~\cite{cho2024jstsp} to convert the waveform into articulatory features. We then calculate the L2 distance between the articulatory features of the current frame and the target phoneme, using the mapping we constructed in Sec.~\ref{sec:art-based}. If the distance exceeds a threshold, denoted as $\tau$, the frame is classified as negative. AER is then computed as the ratio of negative frames to the total length of the speech. In this work, we set $\tau$ to be 0.5 times the distance between the two most distant phonemes.

\begin{table}[th]
    \caption{Phoneme recognition with different training tasks}
    \label{table:eval-loss}
    \centering
    \setlength{\tabcolsep}{4pt} 
    \renewcommand{\arraystretch}{1.2} 
    \resizebox{8cm}{!}{
    \small
    \begin{tabular}{l c c c} 
    \toprule
    Training tasks & PER ($\%$, $\downarrow$) & WPER ($\%$, $\downarrow$) & AER ($\%$, $\downarrow$)\\
    \hline
    \hline
    w/o soft-training & 18.31 & 16.98 & 17.42 \\
    \ \ + smap & 14.81 & 11.98 &  11.87\\
    \ \ + sCTC & 16.14 & 14.36 & 13.29 \\
    \ \ + smap + sCTC & \textbf{12.37} & \textbf{7.41} & \textbf{10.53}  \\
    \bottomrule
    \end{tabular}}
\end{table}

\vspace{-5pt}
\begin{table}[th]
    \caption{Evaluation on real speech datasets}
    \label{table:real-data}
    \centering
    \setlength{\tabcolsep}{5pt} 
    \renewcommand{\arraystretch}{1.2} 
    \resizebox{8cm}{!}{
    \small
    \begin{tabular}{l c c c} 
    \toprule
    Datasets & PER ($\%$, $\downarrow$) & WPER ($\%$, $\downarrow$) & AER ($\%$, $\downarrow$)\\
    \hline
    \hline
    L2-ARCTIC~\cite{zhao18L2Arctic} & 26.67 & 16.53 & 18.79 \\
    Speechocean762~\cite{zhang21speechocean}& 28.33 & 17.74 & 16.33 \\
    MultiPA~\cite{chen24multipa}& 30.49 & 19.01 & 20.45 \\
    PPA Speech~\cite{gorno2011classification-ppa}& 38.63 & 21.78 & 21.45 \\
    \bottomrule
    \end{tabular}}
\end{table}

\vspace{-5pt}
\subsection{Validation}
To assess the performance of the phoneme recognition model, we first conduct evaluation on the VCTK-accent dataset, using three different phoneme similarity modeling methods, and focusing on vowel, consonant, and whole testsets, respectively. The results are presented in Table~\ref{table-eval-vctk}. 
Furthermore, to comprehensively evaluate the impact of phoneme similarity modeling, we trained three additional models with different training objectives: multi-task learning without soft-training, adding soft phoneme mapping training, and adding soft CTC training. For the latter two, the soft training used the articulatory-based method (the best, as referenced in Table~\ref{table-eval-vctk}). We then compared these results with the benchmark (smap + sCTC), as shown in Table~\ref{table:eval-loss}. Additionally, to assess model's generalizability in real-world scenarios, we validated the model on L2-Arctic, Speechocean762, MultiPA, and PPA speech, with the results presented in Table~\ref{table:real-data}.

\subsection{Results and Discussion}
Since there are no established benchmarks suitable for direct comparison, we report PER along with our proposed WPER and AER for each evaluation. As shown in Table~\ref{table-eval-vctk}, the phoneme recognition model's performance on the VCTK-accent testset yields PER consistently around 10–15\%, indicating a solid performance level. Meanwhile, WPER and AER range between 7–12\%, providing a more reasonable measure of the speech pronunciation differences. Notably, vowel substitutions appear to be transcribed more accurately than consonant substitutions, possibly due to the higher TTS simulation quality for vowel substitutions. Among the three phoneme similarity modeling methods, the articulatory-based approach demonstrates the best performance, which aligns with the fundamental principles of human speech production.

From Table~\ref{table:eval-loss}, we can observe that multi-task learning with soft training yields the best results, as it outperforms the other tasks across all three metrics. It reduces the PER from 18.31\% to 12.37\%, and the WPER and AER decrease by approximately 8\%. Among the individual tasks, only adding soft phoneme mapping and soft CTC training show improvements compared to the task without soft training, with soft phoneme mapping demonstrating a greater improvement.
As indicated in Table~\ref{table:real-data}, the model's performance at real speech datasets are not as strong compared to the VCTK-accent test set, but the generalization still at a good level. Among the datasets, the model performs best on L2-Arctic and Speechocean762, which contain read speech texts. Its performance is relatively lower on MultiPA, which features open-response spontaneous speech, and is weakest on PPA speech, representing the most severe form of pronunciation error. 
The performance aligns with the difficulty level of the speech.

%% file: conclusion.tex
\section{Limitation and Conclusion}
This paper presents a framework for verbatim phoneme recognition and error detection using multi-task learning and soft training, incorporating phoneme similarity modeling. Results show that modeling similarity significantly improves transcription accuracy and sets a new benchmark for phonetic error detection. Nonetheless, limitations remain. CMU phonemes may not align well with actual speech; future work could explore IPA~\cite{ipa_wikipedia} or syllables. TTS quality also needs refinement, and LLMs could generate more natural substitution pairs. Finally, integrating articulatory feedback in gestural~\cite{ssdm, lian22bcsnmf, lian2023factor, wu23k_interspeech}, robust visual units~\cite{lian2023av} and kinematic spaces may enable more human-aligned, closed-loop pronunciation learning.

\section{Acknowledgements}
Thanks for support from UC Noyce Initiative, Society of Hellman Fellows, NIH/NIDCD, and the Schwab Innovation fund.